\documentclass[12pt]{article}
\usepackage{amsmath,amssymb,mathrsfs}

\def\Tr{\operatorname{Tr}} 
\def\>{\rangle}\def\<{\langle} \def\sH{\mathscr{H}}
    
\def\mE{\mathcal{E}} \def\aE{\widetilde{\mathcal{E}}}
\def\dual#1{#1^\prime} \def\rank{\textrm{rank}} \def\openone{\pmb{1}}

\begin{document}

\title{On the minimum number of unitaries needed to describe a
  random-unitary channel}

\author{Francesco Buscemi\footnote{email: \texttt{buscemi@qci.jst.go.jp}}\\
  {\small ERATO-SORST Quantum Computation and Information Project,}\\
  {\small  Japan Science and Technology Agency,}\\
  {\small Daini Hongo White Bldg. 201, 5-28-3 Hongo, Bunkyo-ku,
    Tokyo 113-0033, Japan.}}

\date{August 28th, 2006}

\maketitle

\begin{abstract}
  We provide, in an extremely simple way, an upper bound to the
  minimum number of unitary operators describing a general
  random-unitary channel.
\end{abstract}

%%%%%%%%%%%%%%%%%%%%%%%%%%%%%%%%%%%%%%%%%%%%%%%%%%%%%%%%%%%%%%%%%%%%%%

\section{Introduction}

A channel $\mE$---i.~e. a completely positive trace-preserving
map---acting on density matrices $\rho$ defined on a finite
dimensional input Hilbert space $\sH$ (for sake of simplicity, we
consider here channels with equal input and output Hilbert spaces; the
generalization is straightforward) is called \emph{random-unitary} if
it admits a Kraus representation~\cite{kraus} as
\begin{equation}\label{eq:runitary}
  \mE(\rho)=\sum_ip_iU_i\rho U_i^\dag,
\end{equation}
where $p_i$ are probabilities and $U_i$ are unitary operators. This
definition involves an existential quantifier, and there is no known
constructive algorithm to check whether a given channel is
random-unitary or not. Only the necessary condition of being
\emph{unital}, that is, of preserving the identity matrix,
$\mE(\openone)=\openone$, holds\footnote{For two-dimensional systems,
  a channel is random-unitary if and only if it is unital. For higher
  dimensional systems, if a channel is random-unitary it is also
  unital, but the converse does not hold~\cite{landau}.}.

Nonetheless, random-unitary channels play a very special
\emph{physical} role among all possible evolutions that an open
quantum system can undergo~\cite{davies-open}. In fact, Gregoratti and
Werner~\cite{GW} proved that they are the only irreversible channels
that can be perfectly corrected using, as the only side-resource,
classical information extracted from the environment. This property is
actually sufficient and necessary for a channel to be random-unitary
and one would prefer to adopt this one as the \emph{physical and
  operational definition} of random-unitary channels. The idea of
using the environment as a resource then initiated investigations
about \emph{environment-assisted capacities} for quantum
channels~\cite{hayden-king,smolin,winter}. In Ref.~\cite{deco} the
problem also of quantifying the amount of classical information needed
to perfectly correct a random-unitary channel was raised for the first
time in the case of decohering evolutions. In fact, for a given
random-unitary channel, the form~(\ref{eq:runitary}) is highly
non-unique and the Shannon entropy $H(p_i)$ of the probability
distribution weighing the unitaries $U_i$ can be ``artificially'' made
as large as desired. Consequently, in order to derive sensible
information-theoretic relations regarding the information dynamics in
a random-unitary evolution, one has to single out the random-unitary
Kraus representation minimizing $H(p_i)$.

In the present paper we derive an upper bound for the minimal number
of unitary operators needed in Eq.~(\ref{eq:runitary}), thus providing
also a bound to the amount $H(p_i)$ of classical information needed to
be extracted from the environment in order to invert the
random-unitary evolution. Our bound, proved for generic dimension,
does not catch the peculiar geometry that bistochastic qubit channels
enjoy: For $d=2$, it is provably non tight. However, the qubit case is
completely understood and all random-unitary qubit maps have already
been explicitly characterized (see, e.~g. Ref.~\cite{zycz}). In this
sense, a bound for the qubit case is completely useless. On the
contrary, as soon as one leaves the two-dimensional world, already for
$d=3$, the bound we provide is generally non trivial.

\section{Properties of random-unitary channels}

Let us given a channel $\mE$ acting on density matrices $\rho$ defined
on the input Hilbert space $\sH$. As a consequence of the Stinespring
theorem~\cite{stine}, we can write it as follows~\cite{ozawa}
\begin{equation}\label{eq:unitary-real}
\mE(\rho)=\Tr_a[U(\rho\otimes|0\>\<0|_a)U^\dag],
\end{equation}
namely, as a unitary interaction between the system and an
\emph{ancilla} (or \emph{environment}, described by the Hilbert space
$\sH_a$), followed by a trace over the ancillary degrees of freedom.
If the ancilla input state is a pure one---like in
Eq.~(\ref{eq:unitary-real})---Gregoratti and Werner~\cite{GW} proved
that, for all possible unitary interactions $U$ in
Eq.~(\ref{eq:unitary-real}), and for all possible decompositions of
the channel $\mE$ into pure Kraus representations
$\mE(\rho)=\sum_iE_i\rho E_i^\dag$, there exists a suitable rank-one
POVM on the ancilla, let us call it $\{|v_i\>\<v_i|_a\}$,
$\sum_i|v_i\>\<v_i|_a =\openone_a$, such that
\begin{equation}
  E_i\rho E_i^\dag=\Tr_a[U(\rho\otimes|0\>\<0|_a)U^\dag\ (\openone\otimes|v_i\>\<v_i|_a)].
\end{equation}
As an immediate consequence, if the channel $\mE$ admits a
random-unitary decomposition as $\mE(\rho)=\sum_ip_iU_i\rho U_i^\dag$,
with $U_i$ unitary operators, there exists a rank-one POVM on the
ancilla, $\{|\alpha_i\>\<\alpha_i|_a\}$, such that the probability
distribution of its outcomes does not depend on the input state
$\rho$, since
\begin{equation}
  \Tr[U(\rho\otimes|0\>\<0|_a)U^\dag\ (\openone\otimes|\alpha_i\>\<\alpha_i|_a)]=\Tr[p_iU_i\rho U_i^\dag]=p_i,\qquad\forall\rho,\forall i.
\end{equation}

It is now useful to introduce the channel $\aE$ from density matrices
on $\sH$ to density matrices on $\sH_a$ defined as
\begin{equation}
  \aE(\rho)=\Tr_\sH[U(\rho\otimes|0\>\<0|_a)U^\dag].
\end{equation}
Since the unitary interaction $U$ is unique up to local isometries on
$\sH_a$, we can consider such an \emph{ancillary} (or
\emph{complementary}~\cite{holevo}) channel as a canonical one. In
turn, the channel $\aE$ acting on density matrices, induces a unique
\emph{dual} ancillary channel $\dual\aE$ acting on operators $O_a$ on
$\sH_a$ as follows
\begin{equation}
  \Tr[\dual\aE(O_a)\ \rho]=\Tr[O_a\ \aE(\rho)].
\end{equation}
This is nothing but the Heisenberg picture for the ancillary channel
$\aE$. Using this somehow involved notation, we can translate the
Gregoratti and Werner theorem stating that a channel $\mE$ admits a
random-unitary representation~(\ref{eq:runitary}) \emph{if and only
  if} there exists a rank-one POVM $\{|\alpha_i\>\<\alpha_i|_a\}$ such
that
\begin{equation}
\dual\aE(|\alpha_i\>\<\alpha_i|_a)=p_i\openone_a,
\end{equation}
for all $i$, and for some \emph{fixed} probability distribution $p_i$.
In fact
\begin{equation}
  p_i=\Tr[\aE(\rho)\ |\alpha_i\>\<\alpha_i|_a]=\Tr[\rho\ \dual\aE(|\alpha_i\>\<\alpha_i|_a)],\qquad\forall\rho,\forall i.
\end{equation}
In other words, the POVM $\{|\alpha_i\>\<\alpha_i|_a\}$ is mapped to a
classical dice, namely, the POVM $\{p_i\openone_a\}$. Notice that the
cardinality $N$ of the POVM $\{|\alpha_i\>\<\alpha_i|_a\}$ coincides
with the cardinality in the random-unitary
decomposition~(\ref{eq:runitary}).

\section{Extremal rank-one POVM's}

Let us now suppose that $N>(\dim\sH_a)^2$. Then we know that such a
POVM is non extremal~\cite{fuji,partha,mec,nostro,haya} and it can be
convexly decomposed into extremal components
\begin{equation}\label{eq:convex}
  |\alpha_i\>\<\alpha_i|_a=\lambda P_i+(1-\lambda) Q_i.
\end{equation}
(In the above equation we considered a convex combination of just two
extremal terms; the general case does not change the conclusions.)
Since $\{|\alpha_i\>\<\alpha_i|_a\}$ is rank-one and $0<\lambda<1$,
the only possibility to satisfy Eq.~(\ref{eq:convex}) is that the
non-null elements of $\{P_i\}$ and $\{Q_i\}$ are all proportional to
the corresponding element of $\{|\alpha_i\>\<\alpha_i|_a\}$. Hence, by
linearity, also the non-null elements of $\{P_i\}$ and $\{Q_i\}$ are
mapped by $\dual\aE$ to something proportional to $\openone_a$. The
normalization is granted by the normalization of the map $\dual\aE$.
This means that at the end we found two other rank-one POVM's, that is
$\{P_i\}$ and $\{Q_i\}$, that are both extremal, and hence both with
cardinality less or equal to $(\dim\sH_a)^2$, achieving two other
random-unitary Kraus representations for the channel $\mE$.

On the other hand, the normalization condition
$\sum_i|\alpha_i\>\<\alpha_i|_a=\openone_a$ rules out the possibility
that $N<\dim\sH_a$. A von Neumann rank-one measurement, with
$\<\alpha_i|\alpha_j\>=\delta_{ij}$, achieves the lower bound
$N=\dim\sH_a$.

\section{The result}

By now, we showed that a random-unitary channel always admits a
random-unitary decomposition~(\ref{eq:runitary}) involving \emph{at
  most} $(\dim\sH_a)^2$ unitary operators. We can now tighten this
bound by choosing $\dim\sH_a$ as small as possible. The
smallest\footnote{See Ref.~\cite{nostro-real} for a detailed analysis
  of the ancillary space dimension, that is, the ancillary resources,
  needed to implement various possible unitary realizations of a given
  quantum channel.} achievable $\dim\sH_a$ for a given channel $\mE$
coincides with the number of Kraus elements in an
\emph{orthogonal}---or \emph{canonical}---Kraus representation, that
is, $\mE(\rho)=\sum_jK_j\rho K_j$ with $\Tr[K_j^\dag
K_l]\propto\delta_{jl}$. Such a number is precisely the rank of the
Choi-Jamio\l{}kowski~\cite{jam,choi} positive operator $R_\mE$ in
one-to-one correspondence with the channel $\mE$ and defined as
\begin{equation}
R_\mE=(\mE\otimes\mathcal I)|\Omega\>\<\Omega|,
\end{equation}
where $\mathcal{I}$ is the identity channel, and $|\Omega\>$ is a non
normalized ($\|\Omega\|^2=d$) maximally entangled vector in
$\sH\otimes\sH$. An orthogonal Kraus representation of $\mE$
corresponds then to a diagonalization of $R_\mE$.

Thus, we have the main result\bigskip

{\bf Theorem} \emph{A random unitary channel $\mE$ always admits a
  random-unitary Kraus representation}
\begin{equation}
\mE(\rho)=\sum_{i=1}^Kp_iU_i\rho U_i^\dag
\end{equation}
\emph{with}
\begin{equation}\label{eq:bound}
  \rank R_\mE\le K\le(\rank R_\mE)^2.\ \square
\bigskip\end{equation}

The bound~(\ref{eq:bound}) holds regardless of the dimension $d$ of
the input Hilbert space $\sH$. It is then reasonable that it fails in
accurately describing the peculiar case of qubits ($d=2$). In fact, it
is known that \emph{all} bistochastic qubit channels are actually
\emph{Pauli channels} (see, for example, Ref.~\cite{zycz}), that is,
they can always be written as (apart from an overall rotation of the
whole Bloch sphere)
\begin{equation}
\mE(\rho)=\sum_{i=0,x,y,z}p_i\sigma_i\rho\sigma_i,
\end{equation}
where $\{\sigma_0\equiv\openone,\sigma_x,\sigma_y,\sigma_z\}$ are the
usual $2\times 2$ Pauli unitary matrices, and $p_i$ is a probability
distribution. Notice that $\Tr[\sigma_i\sigma_j]\propto\delta_{ij}$:
This means that the Pauli form of qubit bistochastic channels is a
\emph{diagonalization} of the channel itself and the equality
\begin{equation}
K=\rank R_\mE
\end{equation}
holds in this case. However, already for $d=3$ there exist
bistochastic channels that \emph{cannot} be diagonalized on unitary
operators (for an explicit example, see Ref.~\cite{deco}). This
evidence clearly does not prove our bound to be tight. It nonetheless
shows that things, already for $d=3$, acquire highly non-trivial
geometric properties and get much more complicated. In all these
cases, the bound given in the Theorem could be tight.

As an immediate consequence of the Theorem, it stems the
following\bigskip

{\bf Corollary} \emph{The minimum amount of classical information
  needed to be extracted from the environment in order to perfectly
  correct a random-unitary channel $\mE$ is upper bounded as}
\begin{equation}
  H(p_i)\le 2\log(\rank R_\mE).\ \square
  \bigskip\end{equation}
Moreover, since $\rank R_\mE\le d^2$, the following quite loose---yet
independent of the particular channel---bound holds
\begin{equation}
H(p_i)\le 4\log d.
\end{equation}

\section{Concluding remark}

It is noteworthy that we need no more than $(\rank R_\mE)^2$ rank-one
POVM elements in order to extract all the ``useful'' classical
information from the ancilla. This is analogous to what happens in the
case of optimal \emph{accessible information} extraction: as proved by
Davies~\cite{davies-info}, one never needs more that $d^2$ rank-one
POVM elements in order to extract the maximum achievable accessible
information from a $d$-dimensional system. In the case of accessible
information extraction, Davies' bound seems to be tight, in the sense
that examples can be constructed in which the maximum information
gathering is achieved only by a POVM with maximum number of
elements~\cite{shor}. If the analogy is correct, the bound in the
Theorem could be proved to be tight as well, while we expect that the
bound given in the Corollary can be refined.

\emph{Acknowledgments.} This work has been supported by ERATO-SORST
Quantum Computation and Information Project. Discussions with
P.~Perinotti, G.~Chiribella, A.~Winter, and M.~Hayashi are gratefully
acknowledged.

\appendix


\begin{thebibliography}{99}
\bibitem{kraus} K.~Kraus, \emph{States, Effects, and Operations:
    Fundamental Notions in Quantum Theory}, Lect. Notes Phys. {\bf
    190} (Springer-Verlag, Berlin, 1983).
\bibitem{landau} L.~J.~Landau and R.~F.~Streater,
  J.~Lin.~Alg.~Appl.  {\bf 193}, 107 (1993).
\bibitem{davies-open} E.~B.~Davies, \emph{Quantum Theory of Open
    Systems} (Academic Press, London, 1976).
\bibitem{GW} M.~Gregoratti and R.~F.~Werner, J.~Mod.~Opt. {\bf 50},
  915 (2003).
\bibitem{hayden-king} P.~Hayden and C.~King, Quantum~Inf.~Comp., {\bf
    5}, 156 (2005).
\bibitem{smolin} J.~A.~Smolin, F.~Verstraete, A.~Winter, Phys.~Rev.~A
  {\bf 72}, 052317 (2005).
\bibitem{winter} A.~Winter, preprint \texttt{quant-ph/0507045}.
\bibitem{deco} F.~Buscemi, G.~Chiribella, and G.~M.~D'Ariano,
  Phys.~Rev.~Lett. {\bf 95}, 090501 (2005).
\bibitem{zycz} I.~Bengtsson and K.~Zyczkowski, \emph{Geometry of
    Quantum States: an Introduction to Quantum Entanglement}
  (Cambridge University Press, 2006).  See, in particular, Chap.~10,
  Sec.~7.
\bibitem{stine} W.~F.~Stinespring, Proc.~Am.~Math.~Soc. {\bf 6}, 211
  (1955).
\bibitem{ozawa} M.~Ozawa, J.~Math.~Phys. {\bf 25}, 79 (1984).
\bibitem{holevo} A.~S.~Holevo, Prob.~Th.~Appl. {\bf 51}, 133 (2006).
  Available on \texttt{quant-ph/0509101}.
\bibitem{fuji} A.~Fujiwara and H.~Nagaoka, IEEE~Trans.~Inf.~Theory
  {\bf 44}, 1071 (1998).
\bibitem{partha} K.~R.~Parthasarathy, Inf.~Dim.~Anal. {\bf 2}, 557
  (1999).
\bibitem{mec} R.~Mecozzi, Degree Thesis (Pavia, 2002). In Italian.
\bibitem{nostro} G.~M.~D'Ariano, P.~Lo~Presti, and P.~Perinotti,
  J.~Phys.~A:~Math.~Gen. {\bf 38}, 5979 (2005).
\bibitem{haya} M.~Hayashi, \emph{Quantum Information: an Introduction}
  (Springer-Verlag, Berlin, 2006). See Appendix~A.4.
\bibitem{nostro-real} F.~Buscemi, G.~M.~D'Ariano, and M.~F.~Sacchi,
  Phys.~Rev.~A {\bf 68}, 042113 (2003).
\bibitem{jam} A.~Jamio\l{}kowski, Rep. Math. Phys. {\bf 3}, 275
  (1972).
\bibitem{choi} M.-D.~Choi, Lin. Alg. Appl. {\bf 10}, 285 (1975).
\bibitem{davies-info} E.~B.~Davies, IEEE~Trans.~Inf.~Theory {\bf 24},
  596 (1978).
\bibitem{shor} P.~W.~Shor, in \emph{Quantum Communication, Computing,
    and Measurement 3}, ed. by P.~Tombesi and O.~Hirota (Kluwer
  Academic/Plenum Publishers, 2001). Available on
  \texttt{quant-ph/0009077}.
\end{thebibliography}
\end{document}